\documentclass[prb,twocolumn,amsmath,amssymb]{revtex4}

\usepackage{verbatim}
\usepackage[dvips]{epsfig}

\usepackage{amsmath}
\usepackage{amssymb}
\usepackage{wick}

\begin{document}

\title{Combinatorics and Boson normal ordering: A gentle introduction}

\author{P. Blasiak}
\email{pawel.blasiak@ifj.edu.pl}

\author{A. Horzela}
\email{andrzej.horzela@ifj.edu.pl}

\affiliation{H. Niewodnicza\'nski Institute of
Nuclear Physics, Polish Academy of Sciences,\\
ul.\ Eliasza-Radzikowskiego 152, PL 31342 Krak\'ow, Poland}

\author{K. A. Penson}
\email{penson@lptmc.jussieu.fr}

\author{A. I. Solomon}
\email{a.i.solomon@open.ac.uk}

\altaffiliation[Also at ]{The Open University, Physics and Astronomy Department,
Milton Keynes MK7 6AA, United Kingdom}

\affiliation{Laboratoire de Physique Th\'eorique de la Mati\`{e}re Condens\'{e}e,
Universit\'e Pierre et Marie Curie,\\ CNRS UMR 7600,
Tour 24 - 2i\`{e}me \'et., 4 pl.\ Jussieu, F 75252 Paris Cedex 05, France}

\author{G. H. E. Duchamp}
\email{ghed@lipn-univ.paris13.fr}
\affiliation{Institut Galil\'ee, LIPN, CNRS UMR 7030,
99 Av.\ J.-B.\ Clement, F-93430 Villetaneuse, France}


\begin{abstract}
We discuss a general combinatorial framework for operator ordering problems by
applying it to the normal ordering of the powers and exponential
of the boson number operator. The solution of the problem is given in terms of Bell and
Stirling numbers enumerating partitions of a set. This framework reveals several
inherent relations between ordering problems and combinatorial objects, and displays the analytical background to Wick's theorem. 
The methodology can be straightforwardly generalized from the simple example given herein to a wide class of operators.
\end{abstract}

\maketitle

\section{Introduction}
\label{Introduction}

Hilbert space constitutes the arena where quantum phenomena can be described. One common
realization is Fock space, which is generated by the set of orthonormal vectors $|n\rangle$
representing states with a specified numbers of particles or objects. A
particular role in this description is played by the creation $a^\dag$ and annihilation
$a$ operators representing the process of increasing and decreasing the number of
particles in a system, respectively. We consider operators that satisfy the
boson commutation relation $[a,a^\dag]=1$ describing objects obeying Bose-Einstein
statistics, for example, photons or phonons. The fact that the operators $a$ and $a^\dag$ 
do not commute is probably the most prominent characteristic of quantum
theory, and makes it so strange and successful at the same
time.\cite{Dirac,Ballentine}

In this paper we are concerned with the {\em ordering problem} which is one of the
consequences of non-commutativity. This problem derives from the fact that the order in which the
operators occur is relevant, for example, $a^\dag a\neq aa^\dag=a^\dag a+1$. The ordering
problem plays an important role in the construction of quantum mechanical operators. The
physical properties of differently ordered operators may be quite distinct, which we can
see by considering their expectation values. An analysis of operator matrix
elements reveals their physical properties observed as probabilities. There are two sets of states of primary interest in this context:
number states $|n\rangle$ and coherent states $|z\rangle$. The latter, defined as
eigenstates of the annihilation operator $a$, play an important role in
quantum optics\cite{Glauber,Klauder1,Klauder,Louisell,Mandel} and in the phase
space formulation of quantum mechanics.\cite{Schleich}

The calculation of the number or coherent state expectation values reduces to
transforming the original expression to the \emph{normally ordered} form in which all
annihilation operators are to the right. In this form the evaluation of the matrix elements
is immediate. The procedure is called normal
ordering.\cite{Klauder1,Klauder,Louisell, Mandel, Schleich} Although the process is clear and
straightforward, in practice it can be tedious and cumbersome when
the expression is complicated, and is even less tractable when we consider operators
defined by an infinite series expansion. It is thus desirable to find
manageable formulas or guiding principles that lead to solutions of
normal ordering problems.

In this paper we present a general framework that is applicable to a broad class of
ordering problems. It exploits the fact that the coefficients emerging in the normal
ordering procedure appear to be natural numbers which have their origin in combinatorial
analysis. In the simplest case of powers or the exponential of the number operator
$N=a^\dag a$ these are Stirling and Bell numbers which enumerate partitions of a
set.\cite{Katriel} We use this example to illustrate a systematic approach to the
ordering problem. The general methodology is to identify the problem with
combinatorial structures and then resolve it using this identification. The solution may be found
with the help of the {\em Dobi\'nski relation},\cite{Dobinski,Wilf} which is a very effective tool
and is straightforwardly applicable to a wide range of
ordering problems.

As a byproduct of this methodology we obtain a surprising relation between combinatorial
structures and operator ordering procedures. This relation is especially interesting because the
objects involved in the problem can have clear combinatorial interpretations (for example,
as partitions of a set). The expectation is that this remarkable interrelation will
shed light on the ordering problem and clarify the meaning of the associated abstract operator expressions.

The framework we present is an example of a fertile interplay between algebra and
combinatorics in the context of quantum mechanics. It employs only undergraduate
algebra and is not as yet a standard feature of quantum mechanics textbooks.

The paper is organized as follows. Section~\ref{FockSpace} briefly recalls the concept of
Fock space and introduces the normal ordering problem. The main part containing the
connection to combinatorics is given in Sec.~\ref{Generic}. It illustrates the
methodology by discussing in detail the solution of a generic example. 
Some applications are provided in Sec.~\ref{Applications}. In
Sec.~\ref{Summary} we point out extensions of this approach and suggest further
reading.

\section{Occupation number representation}\label{FockSpace}

\subsection{States and operators}

We consider a pair of one mode boson annihilation $a$ and creation $a^\dag$ operators
satisfying the conventional boson commutation relation
\begin{equation}\label{HW}
[a,a^\dag]=1.
\end{equation}
The operators $a, a^{\dag}$, and $1$ generate the Heisenberg algebra.

The occupation number representation arises from the interpretation of $a$ and
$a^\dag$ as operators annihilating and creating a particle in a system. From this point
of view the Hilbert space $\mathcal{H}$ of states is generated by the number
states $|n\rangle$, where $n=0,1,2,\ldots$ counts the number of particles, or objects in
general. The states are assumed to be orthonormal, $\langle
m|n\rangle=\delta_{m,n}$, and constitute a basis in $\mathcal{H}$. This representation is
usually called Fock space.

The operators $a$ and $a^\dag$ satisfying Eq.~(\ref{HW}) may be realized in Fock space as
\begin{equation}\label{an}
a\ |n\rangle = \sqrt{n} |n-1\rangle,\qquad a^\dag\,|n\rangle = \sqrt{n+1}|n+1\rangle.
\end{equation}
The number operator $N$, which counts the number of particles in a system, is defined by
\begin{equation}
N |n\rangle=n |n\rangle,
\end{equation}
and is represented as $N=a^\dag a$. It satisfies the commutation relations
\begin{equation}\label{NN}
[a,N]=a,\qquad [a^\dag,N]=-a^\dag.
\end{equation}
The algebra defined by Eqs.~(\ref{HW}) and (\ref{NN}) describes objects obeying
Bose-Einstein statistics, for example, photons or phonons. It is sometimes called the Heisenberg-Weyl algebra, and occupies a prominent role in
quantum optics, condensed matter physics, and quantum field theory.

The second set of states of interest in Fock space are the {coherent states
$|z\rangle$. They are defined as the eigenstates of the annihilation operator
\begin{equation}\label{az}
a|z\rangle=z|z\rangle,
\end{equation}
where $z$ is a complex number (the dual relation is $\langle z|a^\dag=z^*\langle z|$).
These states take the explicit form
\begin{equation}\label{zn}
|z\rangle=e^{-\frac{1}{2}|z|^2} \sum_{n=0}^\infty\frac{z^n}{\sqrt{n!}}\,|n\rangle.
\end{equation}
These states are normalized, $\langle z|z\rangle=1$, but are not orthogonal and constitute an
overcomplete basis in the Hilbert space.\cite{Overlap} Coherent states have many useful
properties which are exploited in quantum optics
and in other areas of physics.\cite{Glauber,Klauder1,Klauder,Louisell,Mandel,Schleich} 

\subsection{Normal ordering: Introduction}
\label{NormalOrdering}

The noncommutativity of the creation and annihilation operators causes
serious ambiguities in defining operator functions in quantum mechanics. To solve this
problem the order of the operators has to be fixed.
An important practical example of operator ordering is the normally ordered form in which all annihilation operators $a$ stand to the right of the creation operators $a^\dag$.
We now define two procedures on boson expressions yielding a normally ordered form,
namely, normal ordering and the double dot
operation.\cite{Klauder1,Klauder,Louisell,Mandel,Schleich}}

By the normal ordering of a general expression $F(a^{\dag},a)$ we mean
$F^{(n)}(a^{\dag},a)$ which is
obtained by moving all the annihilation operators $a$ to the right using the
commutation relation of Eq.~(\ref{HW}). This procedure yields an
operator whose action is equivalent to the original one, that is,
$F^{(n)}(a^{\dag},a)=F(a^\dag,a)$ as operators, although the form of the expressions in terms of 
$a$ and $a^\dag$ may be completely different.

The double dot operation \mbox{:$F(a^{\dag},a)$:} consists of
applying the same ordering procedure but without taking into account the
commutation relation of Eq.~(\ref{HW}), that is, moving all annihilation operators $a$
to the right as if they commuted with the creation operators $a^\dag$. This notation, although widely used, is not universal.\cite{notation} We observe that in
general this procedure yields a different operator $F(a^{\dag},a) \neq
\mbox{:$F(a^{\dag},a)$:}$ .\cite{remark} 

In addition to the fact that these two procedures yield different results (except for
operators that are already in {normally ordered} form), there is also a practical difference in their use. That is, although the application of the
double dot operation is almost immediate, for the normal ordering procedure some
algebraic manipulation of the non-commuting operators $a$ and $a^\dag$ is needed. Here is an
example of both procedures in action:
\begin{eqnarray}\nonumber
\begin{array}{rcc}
aa^\dag aaa^\dag a&\xrightarrow[\ \ \ \ \ {[a,a^\dag]=1}\ \ \ \ \ ]{\rm normal\ ordering}&
\underbrace{(a^\dag)^2 a^4+4\ a^\dag a^3+2\ a^2}\vspace{2mm}\\
&&\begin{footnotesize}{ a^\dag\ \text{to the left,}\ \ \ a\ \text{to the right}}\end{footnotesize}\vspace{2mm}\\
aa^\dag aaa^\dag a&\xrightarrow[\begin{scriptsize}\begin{array}{c}a,a^\dag\ \text{commute}\\\text{(like numbers)}\end{array}\end{scriptsize}]{\rm double\
dot}&\overbrace{a^\dag a^\dag aaaa}.
\end{array}
\end{eqnarray}
In general we say that the normal ordering problem for $F(a^{\dag},a)$ is solved
if we can find an operator $G(a^{\dag},a)$ for which the following equality is satisfied
\begin{equation}\label{FG}
F(a^{\dag},a) = \mbox{:$G(a^{\dag},a)$:}\ .
\end{equation}
The normally ordered form has the merit of enabling immediate calculation
of an operator's coherent state elements which reduce, by virtue of Eq.~(\ref{az}), to
substituting $a\rightarrow z$ and $a^\dag\rightarrow z^*$ in its functional
representation, that is,
\begin{equation}
\langle z|\mbox{:$G(a^\dag,a)$:}|z\rangle= G(z^*,z).
\end{equation}
Thus, by solving the normal ordering problem of Eq.~({\ref{FG}), we readily obtain
\begin{equation}\label{FGz}
\langle z|F(a^\dag,a)|z\rangle=\ G(z^*,z)\ .
\end{equation}
This procedure may be illustrated in the example
\begin{eqnarray}\nonumber
\langle z| aa^\dag aaa^\dag a\ |z\rangle&=& \langle z| (a^\dag)^2 a^4+4 a^\dag a^3+2 a^2 |z\rangle\\\nonumber
&=&(z^*)^2z^4 +4\,z^*z^3 +2\,z^2 .
\end{eqnarray}
In brief, we have shown that the calculation of coherent state matrix elements reduces to
solving the normal ordering problem. The converse statement is also true; that is, if
we know the coherent state expectation value of the operator, say Eq.~(\ref{FGz}), than
the normally ordered form of the operator is given by Eq.~(\ref{FG}).\cite{Klauder1,Klauder}

A standard approach to the normal ordering problem is to use Wick's theorem.\cite{Wick} In our
context, this theorem expresses the normal ordering of an operator by applying the double dot operation to
the sum of all possible expressions obtained by
removing pairs of annihilation and creation operators where $a$ precedes $a^\dag$, 
called contractions in analogy to quantum field theory, for example
\begin{eqnarray}\nonumber
    &&aa^\dag aaa^\dag a
    =\ :\!\!\!\!\!\underbrace{aa^\dag aaa^\dag a}_{\text{no pair removed}}\!\!\!\!\!:\\\nonumber
    &&+:\underbrace{\wick{1}{<1{\not \!a}>1{\not \!a}\!\!\ ^\dag aaa^\dag a}+\wick{1}{<1{\not \!a}a^\dag aa>1{\not \!a}^\dag a}
+\wick{1}{aa^\dag <1{\not \!a}a>1{\not \!a}\!\!\ ^\dag a}+\wick{1}{aa^\dag a<1{\not \!a}>1{\not
\!a}\!\!\ ^\dag a}}_{\text{1 pair removed}}:\\\nonumber
     &&+:\underbrace{\wick{11}{<1{\not \!a}>1{\not \!a}\!\!\ ^\dag <2{\not \!a}a>2{\not \!a}\!\!\ ^\dag a}+\wick{11}{<1{\not \!a}>1{\not \!a}\!\!\ ^\dag a<2{\not \!a}>2{\not \!a}\!\!\ ^\dag a}}_{\text{2 pairs removed}}:\
    =(a^\dag)^2 a^4+4\ a^\dag a^3+2\ a^2.
\end{eqnarray}
This procedure may involve a large number of steps. For polynomial expressions this difficulty may be
overcome by using computer algebra, although this use does not provide an analytic structure. For nontrivial functions, such as those having
infinite expansions, the problem still remains open.

One approach to the problem relies on the disentangling properties of 
Lie algebraic operators and application of the
Baker-Campbell-Hausdorff formula. 
Here is a standard example:
\begin{equation}
e^{\lambda(a+a^\dag)}= e^{\lambda^2/2}\ \mbox{:$e^{\lambda (a+a^\dag)}$:}\ .
\end{equation}
However, the use of this kind of disentangling property of the exponential operators is 
restricted in practice to quadratic expressions in boson operators.\cite{Wilcox}

Another method exploits the recurrence relations and solves the normal ordering problem by
use of combinatorial identities.\cite{Katriel,KatrielCoherent} This promising approach
was the inspiration for the systematic combinatorial methodology which is presented in
this article.

\section{Generic example: Stirling and Bell numbers}
\label{Generic}

\subsection{Normal ordering: Combinatorial setting}

We consider the number operator $N=a^\dag a$ and seek the normally ordered form of its
$n$th power ($n\geq 1$). We write the latter as
\begin{equation}\label{Snk}
\left(a^\dag a\right)^n=\sum_{k=1}^n S(n,k)\,(a^\dag)^k a^k,
\end{equation}
which uniquely defines the integer sequences $S(n,k)$ for $k=1\ldots n$; these sequences are
called the \emph{Stirling numbers}.\cite{Comtet,Wilf} Information about this
sequence for each $n$ may be captured in the \emph{Bell polynomials}
\begin{equation}\label{B}
B(n,x)=\sum_{k=1}^n S(n,k)\,x^k.
\end{equation}
We also define the \emph{Bell numbers} $B(n)=B(n,1)$ as
\begin{equation}\label{Bn}
B(n)=\sum_{k=1}^n S(n,k).
\end{equation}
Instead of
operators $a$ and $a^\dag$
we may equally well insert into Eq.~(\ref{Snk}) the representation of Eq.~(\ref{HW}) given
by the operator $X$ defined as multiplication by $x$, and by the derivative
$D=\frac{d}{dx}$\cite{XD} 
\begin{equation}
a^\dag\longleftrightarrow X, \qquad 
a\longleftrightarrow D.
\end{equation}
This substitution does not affect the commutator of Eq.~(\ref{HW}), that is, $[D,X]=1$,
which is the only property relevant for the construction. Therefore in this
representation Eq.~(\ref{Snk}) takes the form
\begin{equation}\label{SXD}
\left(X D\right)^n=\sum_{k=1}^n S(n,k)\,X^k D^k.
\end{equation}

\subsection{Combinatorial analysis}

In the following we discuss the properties of the Stirling and Bell
numbers.\cite{Conventions} For that purpose we use elementary methods of combinatorial
analysis based on a versatile tool known as the Dobi\'nski relation.\cite{Dobinski,Wilf} The latter is
obtained by acting with Eq.~(\ref{SXD}) on the exponential function $e^x=\sum_{k=0}^\infty
\frac{x^k}{k!}$ yielding
\begin{equation}
\sum_{k=0}^\infty k^n\frac{x^k}{k!}=e^x\sum_{k=1}^n S(n,k)\,x^k.
\end{equation}
[xx all eqs. must be numbered xx] If we recall the definition of the Bell polynomials Eq.~(\ref{B}), we obtain
\begin{equation}\label{D}
B(n,x)=e^{-x}\sum_{k=0}^\infty \frac{k^n}{k!}\,x^k,
\end{equation}
which is the celebrated \textit{Dobi\'nski relation}.\cite{Dobinski,Wilf} It is usually
expressed in terms of the Bell numbers, which are given by
\begin{equation}\label{DD}
B(n)=e^{-1}\sum_{k=0}^\infty \frac{k^n}{k!}.
\end{equation}
We observe that both series are convergent and express the integers $B(n)$ or
polynomials $B(n,x)$ in a nontrivial way.
To mention one of the many applications notice that $k^n$ in Eq.~(\ref{D}) may be replaced by the
integral representation $k^n=\int_0^\infty d\lambda\, \lambda^n \delta(\lambda-k)$. If we
change the order of the sum and integral (allowable because both are convergent), we
obtain the solution to the Stieltjes moment problem for the sequence of Bell polynomials
\begin{equation}
B(n,x)=\!\int_0^\infty\!d\lambda\ W_x(\lambda)\lambda^n,
\end{equation}
where
\begin{equation}\label{W}
W_x(\lambda)=e^{-x}\sum_{k=0}^\infty \frac{\delta(\lambda-k)}{k!}x^k
\end{equation}
is a positive weight function located at integer points and is called a \emph{Dirac comb}.\cite{BlasiakJPA2006}
Note that Eq.~(\ref{W}) may be identified with the Poisson distribution with mean value equal to $x$.

A very elegant and efficient way of storing and tackling information about sequences is
attained through their generating functions.\cite{Wilf} The {exponential
generating function of the polynomials $B(n,x)$ is defined as
\begin{equation}\label{Gen}
G(\lambda,x)=\sum_{n=0}^\infty B(n,x)\frac{\lambda^n}{n!}.
\end{equation}
It contains all the information about the Bell polynomials. If we use the Dobi\'nski relation,
we may calculate $G(\lambda,x)$ explicitly. We substitute Eq.~(\ref{D}) into Eq.~(\ref{Gen}), change
the summation order,\cite{remark2}
and then identify the expansions of the exponential functions to obtain
\begin{subequations}
\begin{align}
G(\lambda,x)&=e^{-x}\sum_{n=0}^\infty \sum_{k=0}^\infty
k^n\frac{x^k}{k!}\frac{\lambda^n}{n!} \\
&=e^{-x} \sum_{k=0}^\infty \frac{x^k}{k!}\sum_{n=0}^\infty
k^n\frac{\lambda^n}{n!}\\
&=e^{-x} \sum_{k=0}^\infty \frac{x^k}{k!}e^{\lambda
k}=e^{-x} e^{xe^\lambda}.
\end{align}
\end{subequations}
Thus, the exponential generating function $G(\lambda,x)$ takes the compact form
\begin{equation}\label{GFunct}
G(\lambda,x)=e^{x(e^\lambda-1)}.
\end{equation}
Note that in the context of the weight function of Eq.~(\ref{W})
$G(\lambda,x)$ is the moment generating function of the Poisson distribution with the parameter $x$.

An explicit expression for the Stirling numbers $S(n,k)$ may be extracted from the Dobi\'nski relation. Note that in
Eq.~(\ref{D})
the relevant series may be multiplied together using the Cauchy multiplication rule to
yield
\begin{subequations}
\begin{align}
B(n,x)&=\sum_{l=0}^\infty(-1)^l\frac{x^l}{l!} \sum_{k=0}^\infty k^n\frac{x^k}{k!}\\
&=\sum_{k=0}^\infty\sum_{j=1}^k\binom{k}{j}(-1)^{k-j}j^n\frac{x^k}{k!}
\label{26b}\end{align}
\end{subequations}
By comparing the expansion coefficients in Eq.~\eqref{26b} with Eq.~(\ref{B}), we obtain
\begin{equation}\label{Scan}
S(n,k)=\frac{1}{k!}\sum_{j=1}^k\binom{k}{j}(-1)^{k-j}j^n,
\end{equation}
which yields an expression for $S(n,k)$.

If we use any of the above standard formulas Eqs.~(\ref{Scan}) and (\ref{Bn}) for the Stirling or Bell numbers, we can easily calculate
them explicitly. We remark
that many other interesting results may be derived by straightforward manipulation of the
Dobi\'nski relation Eq.(\ref{D}) or the generating function Eq.(\ref{GFunct}), see Appendix
\ref{CombProp}.\cite{BlasiakPhD} Some of these numbers are given in Table:
\begin{eqnarray}
    S(n,k),\ \ \ \ \ \ 1\leq k\leq n\qquad\qquad\qquad\qquad B(n)\ \ \ &&\nonumber\\
    \begin{array}{cl|lllllllllllcc}\cline{3-14}&&&&&&\\
    n=1&&&1&&&&&&&&&&1&\\
        n=2&&&1&1&&&&&&&&&2&\\
        n=3&&&1&3&1&&&&&&&&5&\\
        n=4&&&1&7&6&1&&&&&&&15&\\
        n=5&&&1&15&25&10&1&&&&&&52&\\
    n=6&&&1&31&90&65&15&1&&&&&203&\\
    n=7&&&1&63&301&350&140&21&1&&&&877\\
    n=8&&&1&127&966&1701&1050&266&28&1&&&4140\\
    ...&&&...&...&...&...&...&...&...&...&...&&...
    \end{array}&&\nonumber
\end{eqnarray}

\subsection{Normal ordering: Solution}\label{solution}

We return to normal ordering. By using the properties of coherent states in 
Eqs.~(\ref{FG})--(\ref{FGz}), we conclude from Eqs.~(\ref{Snk}) and (\ref{B}) that the diagonal
coherent state matrix elements generate the Bell polynomials\cite{KatrielCoherent}
\begin{equation}\label{zBz}
\langle z|(a^\dag a)^n|z\rangle=B(n,|z|^2) .
\end{equation}
If we expand the exponential $e^{\lambda a^\dag a}$ and take the diagonal
coherent state matrix element, we find
\begin{equation}
\langle z|e^{\lambda a^\dag a}|z\rangle =\sum_{n=0}^\infty\langle z|(a^\dag
a)^n|z\rangle\frac{\lambda^n}{n!} =\sum_{n=0}^\infty B(n,|z|^2)\frac{\lambda^n}{n!}.
\end{equation}
We observe that the diagonal coherent state matrix elements of $e^{\lambda a^\dag a}$
yield the exponential generating function of the Bell polynomials (see Eqs.~(\ref{Gen})
and (\ref{GFunct}))
\begin{equation}\label{zez}
\langle z|e^{\lambda a^\dag a}|z\rangle=e^{|z|^2(e^\lambda-1)}.
\end{equation}
Equations~(\ref{zBz}) and (\ref{zez}) allow us to read off the normally ordered forms
\begin{align}\label{aa}
(a^\dag a)^n & = \mbox{:$B(n,a^\dag a)$:}\,, \\
\noalign{\noindent and}
\label{eaa}
e^{\lambda a^\dag a} & = \mbox{:$e^{a^\dag a(e^\lambda-1)}$:}\ .
\end{align}
Notice that the normal ordering of the exponential of the number operator $a^\dag a$
amounts to a rescaling of the parameter $\lambda \to e^\lambda-1$. We stress that this rescaling is characteristic for this specific case only and in general the
functional representation may change significantly (see Sec.~\ref{Summary}). Just for illustration we give
results that can be obtained by an analogous calculation\cite{problem}
\begin{align}
\label{21}
e^{\lambda (a^\dag)^2 a} & = \mbox{:$\exp\Big(\frac{\lambda (a^\dag)^2 a}{1-\lambda
a^\dag} \Big)$:}\ ,\\
\label{Kerr}
e^{\lambda (a^\dag)^2 a^2} & = \mbox{:$e^{-a^\dag a}\sum_{n=0}^\infty e^{\lambda n(n-1)}\
\frac{(a^\dag a)^n}{n!}$:}\, .
\end{align}

Equations~(\ref{aa}) and (\ref{eaa}) provide an explicit solution to the normal
ordering problem for powers and the exponential of the number operator. This solution was obtained
by identifying the combinatorial objects and resolving the problem on that basis.
Furthermore, this surprising connection opens a promising approach to the ordering
problem through its combinatorial interpretation.

\subsection{Combinatorial interpretation: Bell and Stirling numbers}
\label{CombInt}

We have defined and investigated the Stirling and Bell numbers as solutions
to the normal ordering problem. These numbers are well known in
combinatorics\cite{Comtet,Wilf} where the $S(n,k)$ are called Stirling numbers of the
second kind. Their original definition is given in terms of partitions of a set; that is, the Stirling numbers $S(n,k)$ count the number of ways of putting $n$ different objects
into $k$ identical containers (none left empty), see
Fig.~\ref{Partitions}. The
Bell numbers $B(n)$ count the number of ways of putting $n$ different objects
into $n$ identical containers (some may be left empty).
From these definitions the recurrence relation of Eq.~(\ref{Rec}) may be readily obtained
and further investigated from a purely combinatorial viewpoint. This formal
correspondence establishes a direct link to the normal ordering problem of the number
operator. As a result we obtain an interesting interpretation of the ordering procedure
in terms of combinatorial objects.
We remark that other pictorial representations can be also given, for example, in terms of
graphs\cite{Mendez} or rook numbers.\cite{QTS3}
\begin{figure}[h]
\hspace{1cm} \resizebox{8.5cm}{!}{\includegraphics{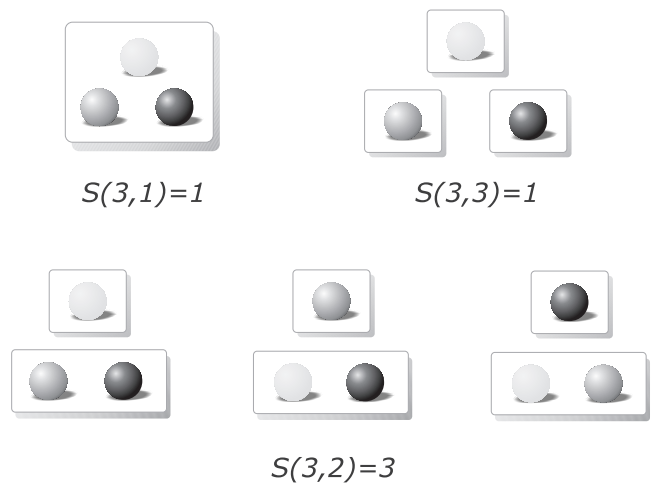}}
\caption{\label{Partitions}
Illustration of Stirling numbers $S(n,k)$ enumerating
partitions of a set of $n=3$ distinguishable marbles (white, gray and black) into
$k=1,2,3$ su
bsets.}
\end{figure}

In conclusion we point out that this method may be reversed; that is, certain combinatorial
families of numbers may be given an algebraic interpretation in the quantum
mechanical context.

\section{Some Applications}
\label{Applications}

\subsection{Quantum phase space}

A curious application of the coherent state representation is found 
in the phase space picture of quantum mechanics. 
For the conjugate pair of observables
${\hat q}={(a^\dag+a)}/{\sqrt{2}}$ and ${\hat p}=i{(a^\dag-a)}/{\sqrt{2}}$,
related to the position and momentum operators of a particle or to quadratures of the
electromagnetic field, coherent state expectation values have the
simple form $\langle z|{\hat q}|z\rangle+i\langle z|{\hat p}|z\rangle=\sqrt{2} z$ and
minimize the uncertainty relation.\cite{Uncertainty} In this sense the coherent state
$|z\rangle$ for $z=(q+ip)/\sqrt{2}$ may be interpreted as the closest quantum
approximation to the classical phase state $(q,p)$. These properties are used to
construct the quantum analog of phase space through the Husimi
distribution, denoted by $Q(q,p)$, which for the quantum state described by the
density matrix $\rho$ is defined as\cite{Ballentine,Schleich}
\begin{equation}\label{Q}
Q(q,p)=\frac{1}{2\pi}\langle z|\rho|z\rangle.
\end{equation}
$Q(q,p)$ is interpreted as the probability density for the system to occupy a region in
phase space of width $\Delta {\hat q}=\Delta {\hat p}=\sqrt{1/2}$ centered at $(q,p)$ which on experimental grounds refers to obtaining the result $(q,p)$ from an optimal simultaneous measurement of ${\hat q}$ and ${\hat p}$. Such measurements in quantum optics are obtained using the technique of heterodyne detection. 

This construction of the quantum phase space analog raises the problem of efficiently calculating the coherent state
expectation values of an operator which, as we have seen in Sec.~\ref{NormalOrdering}, 
is in practice equivalent to its normal ordering. Hence ordering techniques are important for
practical use. 

As an illustration we observe that from Eq.~(\ref{zez}) we can readily derive the explicit expression for the Husimi distribution of the quantum harmonic oscillator in thermal equilibrium. For the Hamiltonian $H=a^\dag a+ 1/2$ the density matrix $\rho$ of a thermal state is $e^{-\beta a^\dag a}/Z$, where $Z=\text{Tr}\,e^{-\beta a^\dag a}=
1/(1 - e^{-\beta})$ and $\beta=1/k_BT$. Thus from Eqs.~(\ref{Q}) and (\ref{zez}) we obtain
\begin{equation}\label{QHO}
Q(q,p)=\frac{1}{2\pi}(1-e^{-\beta})e^{(e^{-\beta}-1)(q^2+p^2)/2}.
\end{equation}
It is instructive to compare this quantum phase space distribution with its classical analog. 
The corresponding Hamiltonian for the classical harmonic oscillator is $H_{\rm cl}=(q^2+p^2)/2$,
and the probability distribution in the thermal state is $P_{\rm cl}(q,p)=e^{-\beta (q^2+p^2)/2}/Z_{\rm cl}$,
where $Z_{\rm cl}=\!\int e^{-\beta (q^2+p^2)/2}\ dqdp=2\pi/\beta$. Finally, we obtain
\begin{equation}\label{PHO}
P_{\rm cl}(q,p)=\frac{1}{2\pi}\beta e^{-\beta(q^2+p^2)/2}.
\end{equation}
In both cases we obtain gaussians. However, observe that the quantum distribution of Eq.~(\ref{QHO}) is wider than its classical analog of Eq.~(\ref{PHO}). It is explained by additional
fluctuations due to the uncertainty relation that are inherent in quantum mechanics. For $\beta\rightarrow 0$, that is, for large temperatures, the quantum distribution of Eq.~(\ref{QHO}) correctly goes to the classical distribution in Eq.~(\ref{PHO}).

An analogous analysis can be done for the whole spectrum of models described by Hamiltonians constructed 
in the second quantization formalism, provided the normally ordered form of the operators is known. 
Section~\ref{Generic} discusses the methodology which is applicable to a wide set of problems, for example, the optical Kerr medium as in Eq.~(\ref{Kerr}) and the open system described by Eq.~(\ref{21}). See Sec.~\ref{Summary} for a discussion of the range of applicability. 

\subsection{Beyond the Wick theorem}

As mentioned in Sec.~\ref{NormalOrdering} the standard approach to normal ordering through Wick's theorem 
reduces the problem to finding all possible contractions in the operator expression. In practice, the 
process may be tedious and cumbersome to perform, especially when a large number of operators are involved. Hence systematic methods, like the one described in Sec.~\ref{Generic}, are of importance in actual applications.

To complete the picture we will show how to connect Wick's approach 
to the combinatorial setting described in this paper. 
The bridge is readily provided by the interpretation of Stirling numbers as partitions of a set, as given in Sec.~\ref{CombInt}. 
To see the link we consider a string $a^\dag a\ a^\dag a\ \ldots\ a^\dag a$ consisting 
of $n$ blocks $a^\dag a$ which we label from $1$ to $n$ starting from the left, 
thus obtaining $n$ distinguishable objects
\begin{equation}\nonumber
\underbrace{a^\dag a}_{1}\ \underbrace{a^\dag a}_{2}\ \ldots\ \underbrace{a^\dag a}_{3}\ \longleftrightarrow\ \textcircled {{\small{\textit{1}}}}\ \textcircled {\small{{\textit{2}}}}\ \, \ldots\, \ \textcircled {\small{{\textit{n}}}}\ .
\end{equation}
Then each choice of contraction in Wick's theorem uniquely divides this set into classes such that objects in the same class are connected by contractions between their operator constituents.
See the following examples for illustration 
\begin{eqnarray}\nonumber
&&\!\!\!\!\!\!\!\!\wick{11}{a^\dag\! <1a\ >1{a}^\dag\!{a}\ {a}^\dag\!a\ {a}^\dag\! <3{a}\ >3{a}^\dag\! a}\ \leftrightarrow\ \ \ \wick{11}{<1{\hbox to-1.2ex{\vbox to.7em{}}}{\kern1.2ex}\ \  \ \  >1 {\hbox to-1.2ex{\vbox to.7em{}}}{\kern1.2ex}\ \ \ \ \ \, {\hbox to-1.2ex{\vbox to.7em{}}}{\kern1.2ex}\ <3 {\hbox to-1.2ex{\vbox to.7em{}}}{\kern1.2ex}\ \ \ \ >3 {\hbox to-1.2ex{\vbox to.7em{}}}{\kern1.2ex}  }\!\!\!\!\!\!\!\!\!\!\!\!\!\!\!\!\!\!\!\!\!\!\!\!\!\!\!\!\!\!\!\!\!\!\textcircled {{\small{\textit{1}}}}\, \textcircled {{\small{\textit{2}}}}\,\textcircled {{\small{\textit{3}}}}\,\textcircled {{\small{\textit{4}}}}\,\textcircled {{\small{\textit{5}}}}\ \leftrightarrow\ \boxed{\textcircled {{\small{\textit{1}}}}\textcircled {{\small{\textit{2}}}}}\ \boxed{\textcircled {{\small{\textit{3}}}}}\ \boxed{\textcircled {{\small{\textit{4}}}}\textcircled {{\small{\textit{5}}}}}\ 
\\\nonumber
&&\!\!\!\!\!\!\!\!\wick{111}{a^\dag\! <1a\ >1{a}^\dag\! <2{a}\ >2{a}^\dag\!a\ {a}^\dag\! <3{a}\ >3{a}^\dag\! a}\ \leftrightarrow\ \ \ \wick{111}{<1{\hbox to-1.2ex{\vbox to.7em{}}}{\kern1.2ex}\ \ \ \,  >1{\hbox to-1.2ex{\vbox to.7em{}}}{\kern1.2ex}\,<2{\hbox to-1.2ex{\vbox to.7em{}}}{\kern1.2ex}\ \ \  >2{\hbox to-1.2ex{\vbox to.7em{}}}{\kern1.2ex}\ \ \ \, <3 {\hbox to-1.2ex{\vbox to.7em{}}}{\kern1.2ex}\ \ \ \, >3 {\hbox to-1.2ex{\vbox to.7em{}}}{\kern1.2ex}  }\!\!\!\!\!\!\!\!\!\!\!\!\!\!\!\!\!\!\!\!\!\!\!\!\!\!\!\!\!\!\!\!\!\textcircled {{\small{\textit{1}}}}\, \textcircled {{\small{\textit{2}}}}\,\textcircled {{\small{\textit{3}}}}\,\textcircled {{\small{\textit{4}}}}\,\textcircled {{\small{\textit{5}}}}\ \leftrightarrow\ \boxed{\textcircled {{\small{\textit{1}}}}\textcircled {{\small{\textit{2}}}} \textcircled {{\small{\textit{3}}}}}\ \boxed{\textcircled {{\small{\textit{4}}}}\textcircled {{\small{\textit{5}}}}}\ 
\\\nonumber
&&\!\!\!\!\!\!\!\!\wick{121}{a^\dag\! <1a\ a^\dag\!<2{a}\ >1{a}^\dag\!a\ >2{a}^\dag\! <3{a}\ >3{a}^\dag\! a}\ \leftrightarrow\ \ \ \wick{121}{<1{\hbox to-1.2ex{\vbox to.7em{}}}{\kern1.2ex}\ \  \ \ <2 {\hbox to-1.2ex{\vbox to.7em{}}}{\kern1.2ex}\ \ \ >1 {\hbox to-1.2ex{\vbox to.7em{}}}{\kern1.2ex}\ \ \ \,>2 {\hbox to-1.2ex{\vbox to.7em{}}}{\kern1.2ex}\,<3 {\hbox to-1.2ex{\vbox to.7em{}}}{\kern1.2ex}\ \ \ >3 {\hbox to-1.2ex{\vbox to.7em{}}}{\kern1.2ex}  }\!\!\!\!\!\!\!\!\!\!\!\!\!\!\!\!\!\!\!\!\!\!\!\!\!\!\!\!\!\!\!\!\!\textcircled {{\small{\textit{1}}}}\, \textcircled {{\small{\textit{2}}}}\,\textcircled {{\small{\textit{3}}}}\,\textcircled {{\small{\textit{4}}}}\,\textcircled {{\small{\textit{5}}}}\ \leftrightarrow\ \boxed{\textcircled {{\small{\textit{1}}}}\textcircled {{\small{\textit{3}}}}}\ \boxed{\textcircled {{\small{\textit{2}}}}\textcircled {{\small{\textit{4}}}}\textcircled {{\small{\textit{5}}}}}\ 
\end{eqnarray}
Observe that this construction may be reversed and thus provides a one-to-one correspondence between operator contractions in $(a^\dag a)^n$ and partitions of the set of $n$ objects.\cite{Bijection}
In this way the contractions of Wick's theorem may be seen as partitions of a set, 
providing the link to the combinatorial framework presented in this paper.
The methodology of Sec.~\ref{Generic} offers an alternative perspective on the normal ordering problem and unlike Wick's approach, exposes its analytical structure, thus yielding practical calculational tools.

\subsection{Operator identities}

Manipulations of differently ordered operators often lead
to interesting operator identities.
For example, taking the limit $\lambda \to -\infty$ in Eq.~(\ref{eaa}), yields an
interesting representation of the vacuum projection operator\cite{Klauder1,Louisell}
\begin{equation}
\label{vacproj}
|0\rangle\langle 0|= \mbox{:$e^{-a^\dag a}$:}\, .
\end{equation}
Equation~\eqref{vacproj} leads to a coordinate representation of the squeezing transformation $S(\lambda) = e^{(\lambda^{*}a^{2} - \lambda a^{\dag 2})/2}$, which is extensively used in quantum optics.\cite{Walls} It may be obtained using the technique of integration within an ordered product,\cite{Fan} yielding
\begin{equation} 
S(\lambda)=e^{-\lambda/2}\!\int\limits_{-\infty}^{\infty}\!dq\, |e^{-\lambda}q\, \rangle\langle\, q|,
\end{equation}
which offers an interpretation of $S(\lambda)$ as an explicit squeezing of the quadrature.

\section{Summary and outlook}\label{Summary}

We have presented a combinatorial framework for operator ordering problems
by discussing a simple example of the powers and exponential of the number
operator $a^\dag a$. We have provided a general method for relating normally ordered
operator expressions to combinatorial objects and then solved the problem from that
viewpoint. The solution involved using the representation of the Heisenberg algebra in
terms of operators on the space of polynomials and then applying the Dobi\'nski relation,
which provides the exponential generating function and explicit expressions. 
This approach provides effective calculational tools and also 
exposes the analytic structure behind the Wick theorem.

One advantage of this methodology is that it can be straightforwardly generalized to a wide class of
operator expressions.\cite{BlasiakPhD} The simplest examples are provided by the powers
and exponentials of $(a^\dag)^ra$ and $(a^\dag)^ra^s$ with $r$ and $s$
integers.\cite{BlasiakPLA} It may be further extended to investigate the normal ordering
of boson monomials\cite{Mendez} in the form
$(a^\dag)^{r_n}a^{s_n}\ldots (a^\dag)^{r_2}a^{s_2}(a^\dag)^{r_1}a^{s_1}$ and more generally
homogeneous boson polynomials,\cite{Gerard} that is, linear combinations of boson
expressions with the same excess of creation over annihilation
operators.\cite{BlasiakPhD} Further development of the method applies to the ordering of
general operators linear only in the annihilation (or creation) operator,
that is, $q(a^\dag) a + v(a^\dag)$, where $q(x)$ and $v(x)$ are arbitrary functions.
The exponential of an operator of this type constitutes a generalized shift
operator and the solution is in the class of Sheffer
polynomials.\cite{BlasiakSheffer} In all of these cases the use of the Dobi\'nski relation
additionally provides a solution of the moment problem,\cite{BlasiakJPA2006} as well as a
wealth of combinatorial identities for sequences involved in the result (including their
deformations\cite{Deformations}).

Ordering problems are naturally inherent in the algebraic structure of quantum mechanics.
It is remarkable that they may be described and investigated using objects having a clear
combinatorial interpretation. For the generic example considered here these are
partitions of a set. For more complicated expressions the interpretation can be provided
by introducing correlations between elements or by using a graph representation.

\appendix

\section{Combinatorial identities}
\label{CombProp}

We enumerate some properties of the Stirling and Bell numbers defined in Sec.~\ref{Generic}.\cite{Abramovitz} The reader is invited to check these relations by
straightforward manipulation of the Dobi\'nski relation or exponential generating
function.\cite{BlasiakPhD}

The recurrence relation for the Stirling numbers is
\begin{equation}\label{Rec}
S(n+1,k)=kS(n,k)+S(n,k-1),
\end{equation}
with the initial conditions $S(n,0)=\delta_{n,0}$ and $S(n,k)=0$ for $k>n$.

The Bell polynomials may be shown to satisfy the following recurrence relation
\begin{equation}
B(n+1,x)=x\sum_{k=0}^n\binom{n}{k}B(k,x),
\end{equation}
with $B(0,x)=1$. Consequently for Bell numbers we have
$B(n+1)=\sum_{k=0}^n\binom{n}{k}B(k)$.

The following exponential generating function of the Stirling
numbers $S(n,k)$ is sometimes used in applications
\begin{equation}\label{GenS}
\sum_{n=k}^\infty S(n,k)\frac{\lambda^n}{n!}=\frac{(e^\lambda-1)^k}{k!}.
\end{equation}
In addition, the Stirling numbers $S(n,k)$ may be interpreted as the connection
coefficients between two sets $x^n$ and $x^{\underline{n}}$, $n=1,2,\ldots$, where
$x^{\underline {n}}=x\cdot(x-1) \ldots (x-n+1)$ is the falling factorial;
that is, they represent a change of basis in the space of polynomials
\begin{equation}
x^n=\sum_{k=1}^n S(n,k)\ x^{\underline {k}} .
\end{equation}

We also note that the Bell polynomials belong to the class of Sheffer
polynomials,\cite{Roman} which in particular share an interesting property called the
Sheffer identity (note the resemblance to the binomial identity)
\begin{equation}
B(n,x+y)=\sum_{k=0}^n\binom{n}{k}B(k,y)\,B(n-k,x) .
\end{equation}


\begin{thebibliography}{999}
\bibitem{Overlap}Coherent states $|z\rangle$ are not orthogonal for different $z$
and the overlapping factor is $\langle
z|z'\rangle=e^{z^*z'-\frac{1}{2}|z|^2-\frac{1}{2}|z'|^2}$. They constitute an
overcomplete basis in the sense of resolution of the identity
$\frac{1}{\pi}\int_\mathbb{C}d^2z\ |z\rangle\langle z|=1$.
\bibitem{notation} The double dot notation is almost universal 
in quantum optics and quantum field theory. Nevertheless some authors, for example, 
Ref.~\onlinecite{Louisell}, use an alternative notation.
\bibitem{remark} Careless use of the double dot notation may lead to inconsistencies, for example if $A=aa^{\dag}$ and $B=a^{\dag}a+1$, we have $A=B$, but :\,$A$\,: $\ne$ :\,$B$\,: . Such problems are eliminated if a rigorous definition, beyond this note is given (to be published by A. I. Solomon, P. Blasiak, G. H. E. Duchamp, A. Horzela, and K. A. Penson).
\bibitem{XD}The simplest representation acts on the space of polynomials and is defined by $X x^n=x^{n+1}$ and $D x^n=nx^{n-1}$. It may be naturally extended to the space of formal power series, see Refs.~\onlinecite{Wilf,Roman}.
\bibitem{Conventions}For convenience and to avoid inaccuracy, the definitions of Stirling and Bell numbers are usually extended by the following conventions: $B(0,x)=B(0)=S(0,0)=1$ and $S(n,k)=0$ for $k=0$ or $k>n>0$.
\bibitem{remark2} Because the generating functions are
formal series, the question of convergence does not arise.
\bibitem{problem} We suggest derivation of these formulas as a problem to be solved by students during classes on the Fock space methods. Detailed calculations may be found in Refs.~\onlinecite{BlasiakPhD,BlasiakPLA}.
\bibitem{Uncertainty}The conjugate operators ${\hat q}$ and ${\hat p}$, satisfying $[{\hat q},{\hat p}]=i$, are subject to the Heisenberg uncertainty relation $\Delta_\psi {\hat q} \Delta_\psi {\hat p}\geq 1/2$. For the coherent state $|z\rangle$ the product of uncertainties exactly equals $1/2$. These are the only states with this property that additionally have equal uncertainties $\Delta_\psi {\hat p}=\Delta_\psi {\hat q}$ (in general, we obtain the squeezed states.\cite{Klauder,Mandel}
\bibitem{Bijection} A detailed proof of this bijection should take into consideration the specific structure of contractions between blocks $a^\dag a$ which make the order in such constructed class irrelevant.
\end{thebibliography}
\end{document}